\documentclass[letterpaper, 10 pt, conference]{ieeeconf}  

\IEEEoverridecommandlockouts                              

\title{\LARGE \bf
Decentralized Merging Control of Connected and Automated Vehicles to Enhance Safety and Energy Efficiency using Control Barrier Functions
}

\author{Shreshta Rajakumar Deshpande$^{1}$ and Mrdjan Jankovic$^{2}$
\thanks{$^{1}$Southwest Research Institute, San Antonio, TX 78238,  USA
        {\tt\small shreshta.rajakumardeshpande@swri.org}}%
\thanks{$^{2}$Southwest Research Institute, Ann Arbor, MI 48108, USA
        {\tt\small mrdjan.jankovic@swri.org}}%
}

\usepackage{graphicx}      

\usepackage{amsmath,amssymb,amsfonts}
\usepackage{algorithmic}
\usepackage{xurl}
\usepackage{textcomp}
\usepackage{xcolor}

\usepackage{amsthm}

\newtheorem{defn}{Definition}

\newcommand{\barr}[2]{\begin{array}{#1}#2\end{array}}

\newcommand{\beq}{\begin{equation}}
\newcommand{\eeq}{\end{equation}}
\newcommand{\mx}[2]{\left[\begin{array}{#1}#2\end{array}\right]}

\newcommand{\R}{\rm{I\kern-2pt R}}

\begin{document}

\maketitle
\thispagestyle{empty}
\pagestyle{empty}

\begin{abstract}

This paper presents a decentralized Control Barrier Function (CBF) based approach for highway merging of Connected and Automated Vehicles (CAVs). In this control algorithm, each ``host" vehicle negotiates with other agents in a control zone of the highway network, and enacts its own action, to perform safe and energy-efficient merge maneuvers. It uses predictor-corrector loops within the robust CBF setting for negotiation and to reconcile disagreements that may arise. There is no explicit order of vehicles and no priority. A notable feature is absence of gridlocks due to instability of the inter-agent system. Results from Monte Carlo simulations show significant improvement in the system-wide energy efficiency and traffic flow compared to a first-in-first-out approach, as well as enhanced robustness of the proposed decentralized controller compared to its centralized counterpart.

\end{abstract}

\section{INTRODUCTION}
Highway merges and interchanges are  major sources of congestion and stress, affecting safety and energy consumption of the transportation system \cite{schrank2015}. Consequently, automation or assistance of merging maneuvers via connected and automated vehicles (CAVs) as well as advanced communication technologies (vehicle-to-vehicle (V2V) and vehicle-to-infrastructure (V2I), for example), has been a subject of significant research. A review of merge-related research can be found in \cite{scarinci} and \cite{rios-torres}. Most of the literature on automated highway merges deal with the one-degree-of freedom (1-DoF) operation where the vehicles are ``on rails" -- that is, a separate system keeps vehicles in the lane, while the merge order is controlled via longitudinal motion planning.

In some merge scenarios there is established priority between vehicles on the main and merge lanes. A typical approach considers requester vehicles in the merging lane and granter vehicles in the highway lane. Slots or gaps would be created by the latter to accommodate merging (\cite{lu, marinescu}). This approach is also considered in the recent International Telecommunication Union (ITU) expert group (EG-ComAD \cite{itu}), a United Nations specialized agency for information and communication technologies. The problem of finding and creating efficient (or even reasonable) slots for all merging vehicles becomes combinatorial when the number of vehicles involved grow. 

In this context, Mixed Integer Programs (MIPs) are introduced to handle the discrete combinatorial component of the merge-related optimization. A model predictive control scheme was combined with a MIP by the authors in \cite{mukai} for the ego (merge road) vehicle. The MIP part selects an appropriate slot to merge -- the order is computed on-line -- while the existence of a sufficient gap is assumed. A MIP was also used in \cite{xu}, where all the vehicles are controlled by a central entity, while the computational complexity is reduced by grouping vehicles into platoons thus lowering the number of options considered by the MIP. When equal priority is assigned to both the main highway and merging lanes, approaches such as first-in-first-out (FIFO) (\cite{xiao2021, xiao2024}) and ``logical ordering" (\cite{kanavalli}) have been used.

In this work, a decentralized 1-DoF controller is proposed to completely avoid establishing explicit vehicle ordering through the merge zone. The algorithm uses control barrier functions (CBFs) (\cite{wieland, ames}) for collision avoidance. In a ``pure" decentralized control, each vehicle plans its own trajectory by using shared traffic information. A distinguishing feature of the proposed decentralized predictor-corrector CBF-based (DPC-CBF) merge controller is that it computes its own actions {\em while considering the constraints and estimating the actions of other agents}. In contrast to pure decentralized algorithms, the predictor-corrector mechanism provides absence of gridlocks due to instability of inter-agent equilibria as discussed in \cite{jankovicTCST, jankovicAR}. This allows us to avoid  ordering or prioritization of the vehicles through the merge zone.

Decentralized CBFs have been previously used in the merge control framework (\cite{liu, xiao2021, xiao2024}) supported by FIFO ordering for gridlock avoidance. The approach discussed in \cite{liu} for on-ramp merging of CAVs in mixed traffic with human-driven vehicles (HDVs) requires V2I communication and the presence of a local coordinator to identify vehicles, specify the merging sequence, and broadcast information to CAVs. In contrast, the method proposed in this work does not require any roadside coordinator. Due to its prevalence in the literature, we used a FIFO algorithm with CBF-based collision avoidance as one of the benchmarks against which the DPC-CBF merging controller is compared.

The authors had previously developed a centralized CBF-based algorithm for 1-DoF merging that also avoids establishing an explicit vehicle order for merging \cite{deshpande2025energy}. In that centralized scheme, each vehicle replicates the central controller computing the velocity controls of all the CAVs within a virtual control zone (CZ). A limitation of the centralized approach is its inability to handle unexpected behavior, especially when vehicles do not follow the commands from the central controller. This scenario may be observed in non-reacting agents and/or vehicles experiencing power loss through a malfunction. Safely responding to such scenarios is a crucial step towards operation in mixed traffic with both CAVs and HDVs. The DPC-CBF merge control algorithm featured in this work is designed to react to others and handle unexpected behavior. Section \ref{sec:unexpected_beh_pwr_loss} in this paper details the capabilities of the DPC-CBF controller in this context.

The contribution of this paper is directed towards allowing human drivers in the mix. First, the algorithm changes from centralized (considered in \cite{deshpande2025energy}) to decentralized. The predictor-corrector CBF algorithm (called PCCA \cite{santillo}) was expected to provide similar behavior between the two, see \cite{jankovicTCST, jankovicAR}, and this is indeed confirmed in the paper. Second, the decentralized algorithm handles unexpected  behavior in stride. Finally, removing the need for the acceleration signal should make  on-board sensing less noise prone in enhancing the situational awareness from BSMs.

\subsection{Assumptions and Requirements}
The situational awareness for operation of the DPC-CBF algorithm is provided by the standard V2V Basic Safety Message (BSM) broadcast by each vehicle at $10$ Hz frequency as specified in the SAE J2735 standard \cite{sae}. The BSM information includes the position, velocity, acceleration, heading, steering angle, size, etc. An advantage of V2V is that it is not sensitive to loss of information due to occlusions that may bring nearby vehicles in or out of the line of sight. The DPC-CBF controllers do not require a central coordinator or V2V range beyond about $75$ to $100$ m. 

\section{PRELIMINARIES: CONTROL BARRIER FUNCTIONS}

In this section, a brief review of robust, exponential control Barrier Functions (CBFs) is provided. Consider the following continuous-time (denoted by subscript $t$) nonlinear system with control-affine dynamics:
\begin{equation} \label{eq:CBF_defn_state_dynamics}
    \begin{aligned}
    \dot{x}_t = f(x_t) + g(x_t) u_t + p(x_t) w_t
    \end{aligned}
\end{equation}
where the vector $x_t \in \mathbb{R}^n$ is the system state, $u_t \in {\mathcal U} \subseteq \mathbb{R}^{n_u}$ is the system input and $w_t \in {\mathcal W} \subseteq \mathbb{R}^{n_w}$ is a bounded external disturbance. Now consider an admissible set ${\mathcal C} = \{ x_t \in \mathbb{R}^n: h(x_t) \ge 0\}$ with the function $h(x_t)$ being differentiable sufficiently many times. Let the notation $L_f h$ denote $\frac{\partial h}{\partial x}f(x_t)$.

\begin{defn}
    The function $h(x_t)$ (compactly notated as $h_t$) is an (exponential) Robust Control Barrier Function (RCBF) if there exists a constant $\lambda$ such that
    \begin{equation} \label{eq:ERCBF_defn}
    \begin{aligned}
    \max_{u_t \in {\mathcal U}}\min_{w_t \in {\mathcal W}}\{ L_f h + L_g h\ u_t + L_p h\ w_t + \lambda h_t\} >0
    \end{aligned}
    \end{equation}
\end{defn}
The definition combines the concepts of an exponential CBF from \cite{nguyen} and RCBF from \cite{jankovicRCBF}. The set $\mathcal C$ is said to be positively invariant as long as the control input satisfying \eqref{eq:ERCBF_defn} is used, $\dot h_t + \lambda h_t >0$ and $h_t > 0, \forall t > t_0$, provided that $x_{t_0}$ is in the admissible set $\mathcal C$.

In this paper, the control inputs and the disturbances will only appear in the second derivative of the CBF $h_t$ i.e., $L_g h = L_p h = 0$. Instead of \eqref{eq:ERCBF_defn}, following \cite{nguyen}, the requirement for $h_t$ to be a second-order CBF is
\begin{equation} \label{eq:ERCBF_defn_2}
   \barr{l} {\max_{u_t\in {\mathcal U}}\min_{w_t\in {\mathcal W}}\{\ddot h_t + l_1 \dot h_t + l_0 h_t \} \\ 
   = \max_{u_t\in {\mathcal U}}\min_{w_t\in {\mathcal W}}\{L_f^2 h + L_g L_f h\ u_t + L_p L_f h\ w_t \\
   \hspace{48mm}  + l_1 L_f h + l_0 h_t\} > 0 }
\end{equation}
where $L^2_f h = L_f(L_f h)$, $l_0 = \lambda_1 \lambda_2$, and $l_1 = \lambda_1 +\lambda_2$. Here, $\lambda_1>0$ and $\lambda_2>0$ are selected such that the two roots of $\chi(s) = s^2 + l_1 s + l_0 = 0$ are negative real numbers. In addition, the invariant set needs to be restricted to ${\mathcal C}_\nu = \{ x_t \in R^n: h_t>0, \lambda_\nu h_t \ge -\dot h_t \}$, where $\lambda_\nu$ is either one of the two roots of $\chi(s)$. It is beneficial to use the smaller one (assumed $\lambda_1$ here) because it makes the admissible set $\mathcal C_1$ larger.

Next, the design of the safety filter for maintaining the state in the safe set $\mathcal C_1$ is reviewed. In \cite{jankovicRCBF}, two options for the disturbance $w_t$ were considered. If the disturbance is unknown, 
 the worst case scenario has to be assumed assumed, and the controller becomes conservative. If we can somehow obtain an estimate of $w_t$, the  controller is less conservative, but the question arises where the estimate comes from. This is addressed in Section \ref{sec:DPC-CBF_algo_details}. 

The standard approach to employ a CBF to guarantee positive invariance of  $\mathcal C_1$ is to override a baseline (or performance) control $u_0$ using a Quadratic Program (QP) setup. With relative degree 2 case and a disturbance estimate $\hat{w}_t \in {\mathcal W}$
assumed available, the QP takes the following form:
\begin{equation} \label{eq:CBF_QP_defn}
\barr{l}{ \min_{u_t \in {\mathcal{U}}}  \|u_t - u_0\|^2 \ \ {\rm such \ that} \\*[2mm]
L_f^2 h + L_g L_f  h\ u_t + L_p L_f  h\ \hat w_t +  l_1 L_f h + l_0 h_t \ge  0  }
\end{equation} 
By definition of the CBF, the solution exists, that is, the QP is feasible.


\section{PROBLEM FORMULATION}
\label{sec:problem_formulation}

\begin{figure}[!t]
	\centering
	\includegraphics[width=\columnwidth]{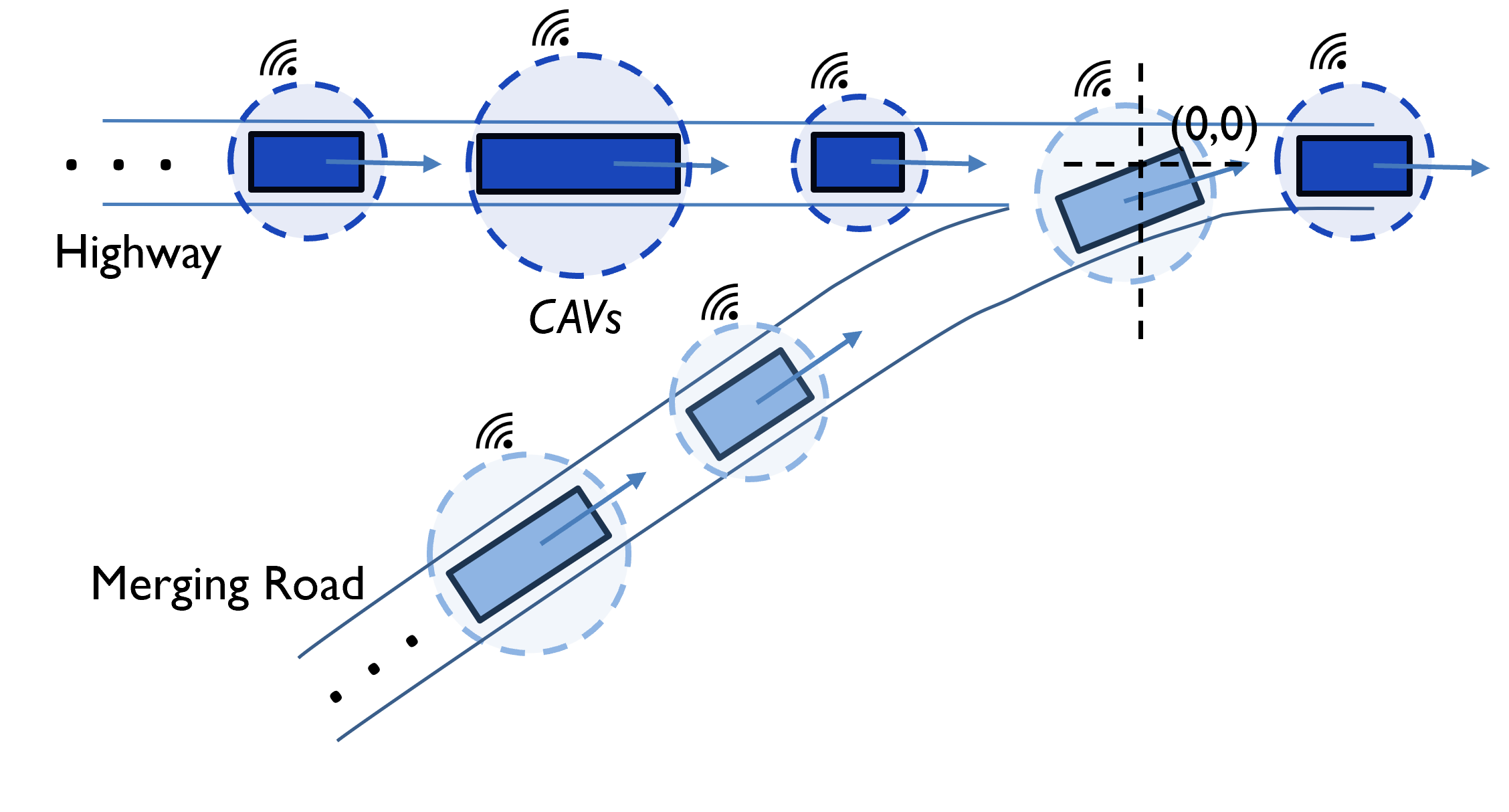}
	\caption{Highway merging scenario with communicating CAVs.}
	\label{fig:short_merge_setup}
\end{figure}

The objective of the proposed algorithm is to control CAVs in a decentralized manner to perform safe merging maneuvers while simultaneously improving their efficiency (in terms of energy consumption and travel time). Fig. \ref{fig:short_merge_setup} illustrates the 1-DoF merge scenario considered, where the merge point is taken as the origin of the coordinate frame. As shown in Fig. \ref{fig:short_merge_setup}, the communicating CAVs that compose the traffic have different sizes (and masses) and may have different desired velocities. Note the absence of a roadside coordinator. In the proposed control scheme, each CAV in a virtual CZ computes and enacts its own optimal action (i.e., decentralized) as well as local copies for others, but their actions computed for each other are not expected to agree. 

\subsection{Modeling and Second-Order Barrier Constraint Formulation} \label{sec:modeling_barrier_constr}
In this work, the vehicles in the traffic network are modeled as double integrators. An agent $i$ is modeled as a disk of radius $r_i$. Note that the proposed approach would also work with other smooth shapes such as ellipses. Its center motion (longitudinal direction only) is given by:
\begin{equation} \label{eq:model_state_eqn}
    \begin{aligned}
    \dot{s}_{t,i} &= v_{t,i} \\
    \dot{v}_{t,i} &= a_{t,i}, \: \text{for all} \: i=1,\dots,N_a
    \end{aligned}
\end{equation}
where $s_{t,i}$ is the distance of vehicle $i$ at time $t$ from the merge point, $v_{t,i}$ is the vehicle velocity, $a_{t,i}$ is its acceleration, and $N_a$ is the number of vehicles in the CZ.

In the DPC-CBF controller model, the longitudinal agent dynamics differs from the plant model in \eqref{eq:model_state_eqn} with the agent's acceleration replaced by a low-pass filter with time constant $\tau_f$ to the agent's velocity control action $u_{t,i}$:
\begin{equation} \label{eq:ctrl_state_eqn}
    \begin{aligned}
    \dot{s}_{t,i} &= v_{t,i} \\
    \dot{v}_{t,i} &= \frac{1}{\tau_f}\left(-v_{t,i} + u_{t,i} \right), \: \text{for all} \: i=1,\dots,N_a
    \end{aligned}
\end{equation}
 The relative motion between CAVs $i$ and $j$ is given by
\begin{equation} \label{eq:rel_motion_eqns}
    \begin{aligned}
    \dot{\xi}_{t,ij} &= v_{t,ij} \\
    \xi_{t,ij} &= \begin{bmatrix} x_{t,i} - x_{t,j} & y_{t,i} - y_{t,j} \end{bmatrix}^\mathsf{T} \\
    v_{t,ij} &= \begin{bmatrix} v_{t,ix} - v_{t,jx} & v_{t,iy} - v_{t,jy} \end{bmatrix}^\mathsf{T}
    \end{aligned}
\end{equation}
where the X-Y location (respectively velocity) of a vehicle $i$, computed from the road geometry, is given by $\{x_{t,i},y_{t,i}\}$ (respectively $\{v_{t,ix},v_{t,iy}\}$); $\xi_{t,ij}$ is the center-to-center (vector) separation between two vehicles and $v_{t,ij}$ is their relative velocity.

The benefits from this filtered velocity formulation are:
\begin{enumerate}
    \item improvement in energy efficiency over the first-order velocity-based approach discussed in \cite{deshpande2025energy},
    \item availability of additional tuning knobs, and 
    \item not requiring acceleration information from other vehicles (the input is included in the BSM but, if a sensor-based backup is desired, acceleration measurements tend to be noisy).
\end{enumerate}
These benefits will be detailed in the sections that follow.

For collision avoidance, the inter-vehicle separation $||\xi_{t,ij}||$ must be greater than the sum of their respective barrier radii, scaled by a tunable safety margin $\beta$ (added for robustness). To achieve this, the barrier function $h_t(\xi_{t,ij})$ is defined by
\begin{equation} \label{eq:barrier_fn}
    \begin{aligned}
    h_t(\xi_{t,ij}) = \xi_{t,ij}^\mathsf{T}\xi_{t,ij} - \left((1+\beta) \left(r_i + r_j \right) \right)^2
    \end{aligned}
\end{equation}

The second-order CBF constraint is then 
\begin{equation} \label{eq:barrier_constraint}
    \begin{aligned}
    F_{t,ij} := \ddot{h}_t + l_1 \dot{h}_t + l_0 h_t = A_{t,ij} + B_{t,ij}(u_{t,i} - u_{t,j}) \geq 0
    \end{aligned}
\end{equation}
where $A_{t,ij} = 2 v_{t,ij}^\mathsf{T}v_{t,ij} + 2\xi_{t,ij}^\mathsf{T}v_{t,ij}\left(l_1 - \frac{1}{\tau_f}\right) + l_0h_t(\xi_{t,ij})$ and $B_{t,ij} = \frac{2}{\tau_f}(\xi_{t,ij})^\mathsf{T}$. One CBF constraint is generated per pair of vehicles i,e., the total number of barrier constraints in the CZ is $\frac{N_a(N_a-1)}{2}$.

\section{DECENTRALIZED CBF MERGING CONTROL ALGORITHM} \label{sec:DPC-CBF_algo_details}

The decentralized predictor-corrector CBF-based controller presented here is not one of the standard decentralized methods (``decentralized-follower" if other vehicles may be non-responding or ``decentralized-reciprocal" if they are known to be able to take its share of responsibility to avoid collisions, see \cite{jankovicTCST}). In the proposed DPC-CBF control algorithm, each ``host" vehicle negotiates with other agents to perform merge maneuvers in a safe and energy-efficient manner. It does so by computing the optimal action (estimate) for all the agents and then using the error between the computed control estimates for the other agents and the corresponding values  from observations or received via BSMs. These errors are treated as known disturbances and utilized in the CBF-based algorithm. This mechanism is called predictor-corrector CBF algorithm and the method, first introduced in \cite{santillo}, as PCCA.

The host vehicle $i$ computes the controls $u_{t,j|i}$ i.e., the control of agent $j$ estimated by $i$, and implements its own optimal action $u_t^i=u_{t,i|i}^*$. The following QP is formulated for host $i$:
\begin{equation} \label{eq:decentralized_PCCA_cost_fn_constr}
    \barr{l}{ \min_{u_{t,1|i},\dots,u_{t,N_a|i}} \left\Vert u_{t,i|i} - v_{t0,i}\right\Vert^2 + \bar m_{v,i} \left\Vert a_{t,i} \right\Vert^2 \\
    \hspace{15mm} + \sum _{j=1,j \neq i}^{N_a} \left\Vert u_{t,j|i} - v_{t,j}\right\Vert^2 + \bar m_{v,j} \left\Vert a_{t,j|i} \right\Vert^2\\
    \text{subject to } A_{t,ij} + B_{t,ij}\left(u_{t,i|i} - u_{t,j|i} - \hat{w}_{t,j|i} \right) \geq 0,\\
    \hspace{40mm} \text{for all} \: j = 1,\dots, N_a, \: i\neq j \\
    \hspace{2mm} A_{t,jp} + B_{t,jp}\left(u_{t,j|i} + \hat{w}_{t,j|i} - u_{t,p|i} - \hat{w}_{t,p|i} \right) \geq 0,\\
    \hspace{23.5mm} \text{for all} \: j,p = 1,\dots, N_a, j\neq p, \: j,p\neq i }
\end{equation}
where $v_{t0,i}$ is the desired velocity of the host $i$, and $\bar m_v = \alpha\tau_f^2 m_{v}$ denotes its scaled mass ($m_{v}$ being the actual vehicle mass and $\alpha$ a scaling factor) -- heterogeneous traffic is accounted for by introducing a penalty on vehicle acceleration scaled by the vehicle mass. The desired velocity for the host $i$ may be defined as a fixed setpoint or a time-varying reference from a cruise control feature. The acceleration term is computed using \eqref{eq:ctrl_state_eqn}: $a_t = \frac{1}{\tau_f}\left(-v_t + u_t \right)$. In this work, $v_{t,j}$, the actual velocity of agent $j$ at time $t$, is used as an estimate of its desired velocity. In other words, since desired velocities for other agents are not known, their current velocities are used as desired.

A special feature in this decentralized formulation is the addition of the $\hat{w}_t$ terms in the constraints, referred to as disturbance estimates. These terms, used for negotiations between agents, are computed as the difference between the optimal control action for agent $j$ estimated by the host $i$ ($u_{t,j|i}^*$) with the action agent $j$ actually implemented ($u_{t,j}$):
\begin{equation} \label{eq:w_hat_defn}
    \begin{aligned}
    \hat{w}_{t,j|i} = u_{t,j} - u_{t,j|i}^*
    \end{aligned}
\end{equation}
Clearly, $\hat{w}_{t,i|i} = 0$. Here, $u_{t,j|i}^*$ requires knowledge of $\hat{w}_{t,j|i}$ and vice versa, creating an algebraic feedback loop. This is overcome by implementing the disturbance update as a low-pass filter with time constant $\tau_{w}$:
\begin{equation} \label{eq:w_hat_update_discrete_time}
    \begin{aligned}
    \dot{\hat{w}}_{t,j|i} = \frac{1}{\tau_w} (-\hat{w}_{t,j|i} + u_{t,j} - u_{t,j|i}^* ), \: \forall \: i,j=1,\dots,N_a
    \end{aligned}
\end{equation}
In the decentralized scheme, $u_{t,j}$ is not known directly  and is estimated using \eqref{eq:ctrl_state_eqn}, where the vehicle velocity and acceleration are obtained via on-board sensors or BSM messages. Alternately, by selecting $\tau_w = \tau_f = \tau$, $u_{t,j}$ and $u_{t,j|i}^*$ can be separately filtered as: $u_{t,j}^{filt} = \frac{1}{\tau s + 1}u_{t,j}$ and $u_{t,j|i}^{*,filt} = \frac{1}{\tau s + 1} u_{t,j|i}^*$.
After a brief initial transient $u_{t,j}^{filt} = v_{t,j}$ and then, $\hat{w}_{t,j|i} = v_{t,j} -  u_{t,j|i}^{*,filt}$ is computed, thereby eliminating the need for acceleration data from other vehicles. This is a major benefit of the filtered velocity formulation compared to the standard acceleration-based approach. 


The collision constraints are augmented to include additional box constraints. Here, the following limits were applied on the controller for host vehicle $i$ to constrain its acceleration: $\frac{1}{\tau_f}(-v_{t,i} + u_{t,i} ) \in [-6, 5]$ m/s$^2$. 

The described optimization problem \eqref{eq:decentralized_PCCA_cost_fn_constr} was solved using qpOASES, a structure-exploiting active-set QP solver. Note that this is solved at discrete time instances for practical applications. The worst-case QP compute time in each host vehicle per step for the 20-agent simulations is just $12$ ms on a laptop\footnote{$2.5$ GHz $13^{\text{th}}$ Gen Intel Core i7-13800H, $16$ GB RAM}, verifying its computational viability.

A key feature of this DPC-CBF algorithm is that it can be integrated with existing longitudinal velocity controllers as a safety filter, including adaptive cruise control (ACC) systems as well as advanced Eco-driving controls \cite{deshpande2022real,wang2023connected,deshpande2024real}.

\subsection{Four-Vehicle Simulation: An Example} \label{sec:DPC-CBF_example_4veh_sim}
In this section, an example simulation is used to illustrate the functionality of the DPC-CBF algorithm. The defined CZ comprises road sections before and after the merge point, having lengths of $200$ m and $350$ m respectively. The section beyond the merge point accounts for re-acceleration after the merge maneuver for energy efficiency calculations. Here, a challenging scenario was set up, having nearly symmetric initial conditions and four vehicles in the CZ -- two each on the highway (IDs \{H1,H2\}) and merging lanes (IDs \{M1,M2\}). These vehicles have the same initial (and desired) velocities ($20$ m/s) and the same masses ($4500$ lbs). The highway agents are separated by a time gap of $2$ s; the first merge-road agent M1 (closest to the merge point) is just ahead of the first highway agent H1 by $0.1$ m (in terms of distance to the merge point) and the second merging agent M2 is just behind H2 by $0.1$ m. The sampling time is $T_s = 0.1$ s, which is consistent with the BSM broadcast rate.

The DPC-CBF control parameters are $\{\lambda_1, \lambda_2\, \tau_f, \tau_w \}  = \{0.6, 2.0, 0.4, 0.4 \}$. As described in Section \ref{sec:DPC-CBF_algo_details}, setting $\tau_f = \tau_w$, removes the need for acceleration data from surrounding vehicles. It is worth noting, however, that selection of $\tau_w$ has collision avoidance impact.: slower filter time constant $\tau_w$ may require larger barrier margin $\beta$.

\begin{figure}[!t]
	\centering
	\includegraphics[width=\columnwidth]{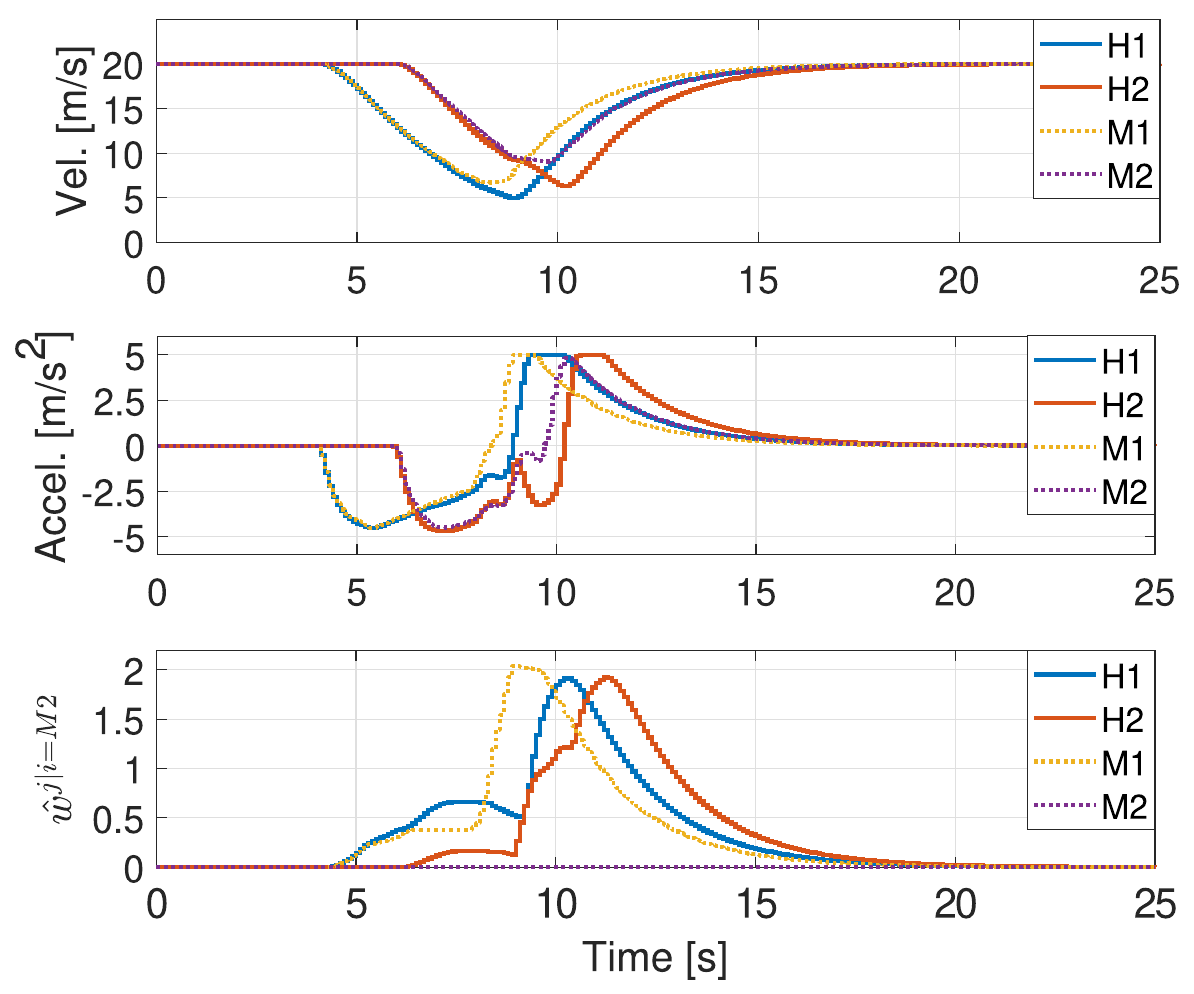}
	\caption{Simulation with four agents showing velocity, acceleration and disturbance estimate traces.}
	\label{fig:DPC-CBF_4agent_scenario}
\end{figure}

Fig. \ref{fig:DPC-CBF_4agent_scenario} shows the results from this simulation. The DPC-CBF controller results in fast negotiation between the vehicles and while assuring collision avoidance. Upon activation of the barrier constraints, the vehicles decelerate to execute the merging maneuver. Though the initial conditions are nearly symmetric, the vehicles do not slow down below $5$ m/s. The control parameters were tuned such that the decelerations remain smooth and keep a sufficient margin from the $-6$ m/s$^2$ limit. One of the underlying disturbance estimate terms in the DPC-CBF algorithm, computed by vehicle M2 on the merging lane i.e. $\hat{w}_{j|i=M2}$, is shown in Fig. \ref{fig:DPC-CBF_4agent_scenario}. This term reconciles differences that may arise between the controls estimated by the host M2 and the actions actually implemented by the other vehicles (note that $\hat{w}_{j=M2|i=M2}=0$, as expected). The final merging order was observed to be M1-H1-M2-H2 while the merging order from a standard first-in-first-out controller would be M1-H1-H2-M2. This highlights a distinguishing feature of the proposed DPC-CBF approach -- it is an unstructured algorithm that does not provide an established precedence or passing order through the merge zone.

\subsection{Tuning the Parameters} \label{sec:DPC-CBF_feasibility_instability}
The addition of tuning parameters in the filtered-velocity approach has 
brought up a question of how to tune them. Here, for clarity, we consider a case with only two vehicles, one in each lane. Only the centralized controller is considered -- the controller has full information and does not use predictor-corrector loops. That is, instead of the current velocities for the other agents in
(\ref{eq:decentralized_PCCA_cost_fn_constr}), their actual desired velocity $v_{t0,i}$ is used.

If their mutual constraint does not activate, the control action for each agent leads to 
\[ \dot v_{t,i} = -\frac{\tau_f}{\tau_f^2 + \bar m_v}(v_{t,i} - v_{t0,i}), \ \ i = 1,2\]
Obviously, $\bar m_v$ attenuates the response of the velocity filter -- the heavier the vehicle, the slower its response.

Next, for the contested case when the vehicles have to negotiate their passing order, i.e. their agent-to-agent constraint activates, the control actions are given by 
\[ \mx{c}{u_{t,1}^* \\ u_{t,2}^*} =  \mx{c}{\bar v_{t,1} \\ \bar v_{t,2} } - \frac{a  + b_1 \bar v_{t,1} + b_2 \bar v_{t,2}} { b_1^2 + b_2^2} \mx{c}{b_1 \\ b_2} \]
where $a$, $b_1$ and $b_2$ are scalars that correspond to elements of the $A$ and $B$ matrices, following the notation in (\ref{eq:decentralized_PCCA_cost_fn_constr}), $\bar v_{t,1} = \frac{v_{t0,1} +v_{t,1} \bar m_v/\tau_f^2}{1+ \bar m_v /\tau_f^2}  $, and $\bar v_{t,2} = \frac{v_{t0,2} + v_{t,2} \bar m_v/\tau_f^2}{1+ \bar m_v /\tau_f^2}  $.

The subsequent analysis is much easier to carry out in transformed coordinates in which the agent positions and velocities appear orthogonal:
\[ z = T s_t, \ v^z = Tv_t, \ {\rm where} \ \  T = \mx{cc}{ 1 & -\cos \gamma \\
0 &\sin \gamma}\]
and $\gamma$ is the angle between the two lanes. 
Note that the change of coordinates does not impact the subsequent stability analysis since it does not change the eigenvalues of the state matrix of the linearized system. Moreover, the inverse transformation $T^{-1}$ can be applied to return to the original coordinates. Thus, in the new coordinate system the equilibrium point (notated by subscript $e$) is defined by setting the right hand side of the $(z, v^z)$ dynamics to 0 to obtain
\[ h_e(z) = z_e^T z_e - D^2, \ z_{1e} v^z_{t0,2} = z_{2e} v^z_{t0,1}\]
where $D = 2 (1+\beta) r$ (agents of the same size are assumed) and $z_e = [z_{1e}, \ z_{2e}]^\mathsf{T}$. 

Linearized system has 2 separate groups of states: (1) the 
$\eta = [z_{1e}, \ z_{2e}] z$, $v^\eta = [z_{1e}, \ z_{2e}] v^z$
state matrix has eigenvalues inherited from the $h$-dynamics: $\{-\lambda_1, -\lambda_2\}$; (2) the $\mu = [z_{2e}, \ -z_{1e}] z$, $v^\mu = [z_{2e}, \ -z_{1e}] v^z$ state matrix is
\[ A^\mu = \mx{cc}{ 0 & 1 \\ \frac{\kappa\|v_{t0}\|}{D} & -\kappa }\]
where $v_{t0} = [v_{t0,1}, \ v_{t0,2}]^\mathsf{T}$ and $\kappa = \frac{\tau_f}{ \tau_f^2 + \bar m_v}$. For the simulations, we have used $\bar m_v = \alpha m_v \tau_f^2$ and $\kappa = \frac{1}{ \tau_f(1+  \alpha m_v)}$. The eigenvalues of $A^\mu$ are $\{ -\kappa/2 \pm \sqrt{\kappa^2/4+\frac{\kappa\|v_{t0}\|}{D} } \}$. Note that one of the two eigenvalues is positive and the $\mu$-dynamics is unstable. 

In contrast to \cite{jankovicTCST, jankovicAR}, where the unstable eigenvalue couldn't be adjusted, there exists some control over it here, but the range is fairly limited. A standard controls root-locus argument for the transfer function $G(s) = \frac{s -\frac{\|v_{t0}\|}{D}}{s^2}$  shows that the possible range of the unstable eigenvalue is $(0, \frac{\|v_{t0}\|}{D})$ as $\kappa$ goes from 0 to  $\infty$. For performance, i.e. fast negotiation, one would like the system as unstable as possible. On the other hand, if $ \frac{\|v_{t0}\|}{D}$ is large relative to the sampling rate, one could use $\kappa$ to attenuate the instability. In the simulations, for the average size and speed of the vehciles and selected tuning, the unstable eigenvalue is $1.7s^{-1}$.

\section{EVALUATION AND RESULTS}
\label{sec:evaluation_results}

\subsection{Benchmark Controllers}
The DPC-CBF merge control algorithm was evaluated against two benchmarks: a FIFO scheme and a centralized CBF-based merge controller.

\subsubsection{FIFO Control}
In the FIFO control method, the first vehicle to enter the defined CZ in is the first one out. Assignment of the merging priority typically requires a central coordinator. The FIFO approach is itself quite efficient than human drivers and likely results in better traffic flow. It is often used as a component of several intersection and merge control algorithms.

In this work, each FIFO vehicle was modeled as a double integrator with the acceleration as the control input. A second-order CBF constraint was used for collision avoidance. The constraints were formulated for vehicles having higher priority, while lower priority ones (i.e., those behind it) were ignored. The same acceleration limits were applied in the FIFO case (in addition to the priority-based barrier constraints for safety). Additionally, a slack variable was used here to guarantee feasibility. 

\subsubsection{Centralized-CBF Merging Control}
In the Centralized-CBF (C-CBF) approach, a central control scheme optimizes for all the CAVs in the CZ. A requirement of the C-CBF approach is the exact knowledge of the current desired speeds of all the CAVs (assumed available via {\em augmented} BSMs as proposed in \cite{deshpande2025energy}).

While the formulation described in Section \ref{sec:problem_formulation} is otherwise retained, the centralized architecture does not compute feedback-based disturbance estimates. Teh lack of any reconciliation mechanist makes it less robust than  DPC-CBF to unexpected (driving) behavior. Note that in the absence of such behavior i.e., ``normal" operation with all vehicles responding fully as planned, the C-CBF control assures collision avoidance while merging, see \cite{deshpande2025energy}.

\subsection{Monte Carlo Simulation Setup} \label{sec:MC_sim_setup}
Evaluation of the DPC-CBF merging controller was performed using Monte Carlo simulations, where $20$ vehicles were considered -- $10$ each on the highway and merging roads. The CZ geometry was specified in Section \ref{sec:DPC-CBF_example_4veh_sim} while the simulation setup is summarized in Table \ref{table:MC_sim_setup}.

The initial (and desired) speeds for the CAVs were  uniformly randomly distributed between $20$-$25$ m/s. The vehicle injection rate was also randomly selected (uniform distribution) between $1100$-$1200$ vehicles per hour per road. The chosen $m_{base}$ was $2375$ lbs and corresponds to a Mitsubishi Mirage while the largest vehicle mass corresponds to a Chevrolet Silverado EV (four times heavier). The vehicle masses were selected randomly, having uniform distribution within these two extremes. The target coast-down coefficients for these vehicles were obtained from the EPA \cite{EPAtestdata}. Having a range of $2$-$4$ m, the barrier radius of each vehicle was computed via linear scaling based on the selected $m_{v}$.

For the DPC-CBF, the same control parameters and the sampling time $T_s$ chosen for the example in Section \ref{sec:DPC-CBF_example_4veh_sim} were used here. The same $\{\lambda_1, \lambda_2 \}$ values were chosen in the C-CBF algorithm as well, while the FIFO used $\{\lambda_1, \lambda_2, M \} = \{0.3, 2.0, 10^4 \}$. Here, $M$ is a tuning weight that scales the slack variable in the cost function of the FIFO controller. The DPC-CBF and C-CBF were run with hard constraints (i.e. no slack variables).

\begin{table}[!t]
\centering
\caption{Setup for Monte Carlo simulations.}
\label{table:MC_sim_setup}
\begin{tabular}{cc}
\hline
\textbf{Parameter or Variable} & \textbf{Value} \\ \hline
Merge road angle & $30^\circ$  \\
Initial (desired) velocity & $v_{k0} \sim \mathcal{U}\left(20, 25\right)$ m/s  \\
Injection rate (per road) & $\sim \mathcal{U}\left(1100, 1200\right)$ veh/hr \\
Sampling time & $T_s = 0.1$ s \\
Vehicle mass & $m_v \sim \mathcal{U}\left(m_{base}, 4 \cdot m_{base}\right)$ lbs \\
Vehicle (barrier) radius & $2-4$ m \\ 
Barrier margin & $\beta = 10\%$
\end{tabular}
\end{table}

\subsection{Robustness to Unexpected Behavior (Power Loss Due to Malfunction)} \label{sec:unexpected_beh_pwr_loss}

Having the capability to handle unexpected behavior encountered while driving is crucial towards real-world applications, where certain vehicles may not respond as predicted. One such scenario is when a vehicle experiences a malfunction that results in power loss. The vehicle would start decelerating at road-load rate and coast down to a standstill. This scenario is used to compare the proposed DPC-CBF algorithm against the C-CBF benchmark, highlighting one of the main benefits of the proposed architecture.

Monte Carlo simulations were performed to compare the robustness of the two algorithms. In total, 100 simulations were carried out -- in 50 of them, one of the highway agents experienced power loss while in the other 50, a merge agent experienced power loss. These agents were chosen to be among the middle of the pack, thereby presenting a challenging scenario for evaluation. For fair comparison, the same initial conditions and vehicle masses were used in the C-CBF and DPC-CBF runs. Other details of the simulation setup follow Section \ref{sec:MC_sim_setup}. Note that in this scenario, the vehicles are fully automated and are not equipped with additional collision avoidance systems, other than the corresponding CBF-based safety filter. 

\begin{figure}[!t]
	\centering
	\includegraphics[width=0.85\columnwidth]{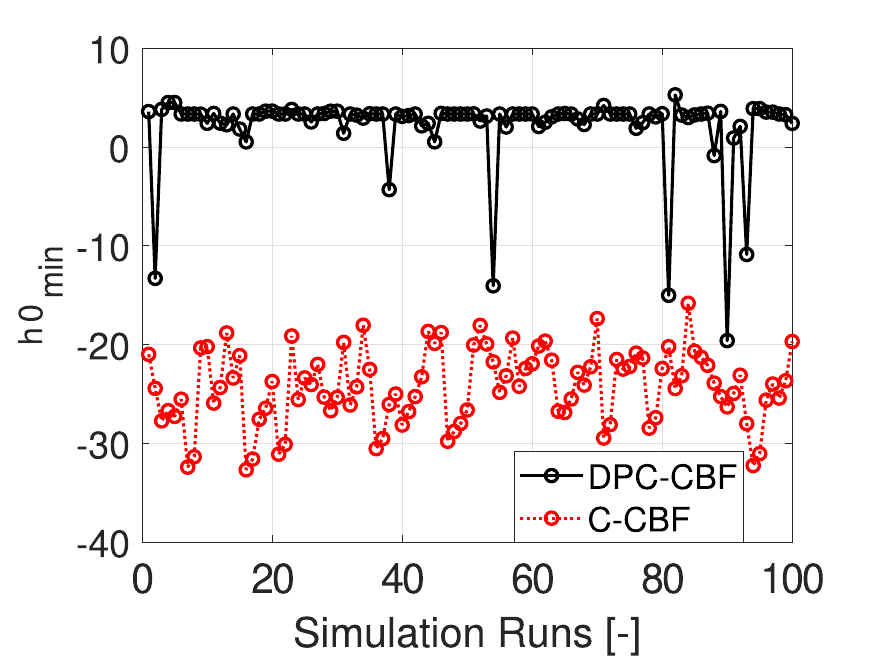}
	\caption{Robustness to unexpected behavior: Comparison of minimum barrier distance.}
	\label{fig:h0min_comp_eval_pwr_loss}
\end{figure}

The main metric used for evaluation is the minimum inter-vehicle barrier distance $h0_{min}$. Computed for all agents during each simulation run, $h0_{min}$ corresponds to the barrier safety margin $\beta = 0$ in \eqref{eq:barrier_fn}, rather than $\beta = 0.1$ used by the DPC-CBF and C-CBF controllers in simulations. When $h0_{min} = 0$, the barrier circles of two vehicles touch and  $h0_{min} < 0$ indicates a collision may occur between these vehicles. As seen in Fig. \ref{fig:h0min_comp_eval_pwr_loss}, all runs for the C-CBF case result in collisions i.e., necessitating takeover by other collision avoidance systems, which is abrupt, inefficient, and may still not guarantee safety. In some cases, the CAVs preceding the malfunctioning vehicle expect it to: (1) either follow its desired speed or (2) in many cases, accelerate prior to or just ahead of the merge point to safely avoid surrounding vehicles while executing the merging maneuver. In other cases, the CAVs ahead of the malfunctioning vehicle may expect it to slow down at a rate greater than its current deceleration. In both these scenarios, the centralized controller, without having predictor-corrector terms in its formulation, cannot accommodate such departures from expected behavior.

In contrast, collisions occur in only 7 of the 100 runs with the DPC-CBF controller. That is, use of the DPC-CBF algorithm saw reductions in collisions by over $90\%$. Overall, these simulations highlight the robustness of the DPC-CBF formulation, which is promising and essential towards real-world operation alongside human driven vehicles.

\subsection{Metrics for Merging Energy and Time Efficiency Evaluation}
In this work, travel or merging time is defined as the time taken by the last vehicle in the simulation to cross the merge point. Evaluation of the DPC-CBF algorithm was performed against the FIFO and C-CBF benchmarks based on flow-related metrics (i.e., travel time and average vehicle speed) as well as the following energy consumption metrics: Positive acceleration Kinetic Energy (PaKE), Braking Energy (BE) and Total Energy Loss (TEL). Equations that define these powertrain-agnostic metrics have been detailed in \cite{deshpande2025energy}.

The PaKE of a vehicle at each discrete time instant, a measure of energy consumed in (re)accelerations, is given by the difference of squared velocity between successive discrete time steps, multiplied by the vehicle mass. This is computed only when the acceleration is positive. The BE of a vehicle accounts for the braking energy in excess of the energy from coasting-based deceleration. This requires computation of the road load losses, based on the vehicle-specific dyno coefficients published by the U.S. EPA \cite{EPAtestdata} for each vehicle they test. The TEL of a vehicle computes the maximum between the energy from braking action and the road load energy loss, and captures the total energy losses. Each of these metrics are normalized by the distance traveled. Note that these definitions provide vehicle-level energy-related metrics; the corresponding system-level metrics are obtained by averaging them across all the $N_a$ vehicles in the CZ.

\subsection{Evaluation of Merging Energy and Time Efficiency} \label{sec:MC_sim_energy_time_eval}
To evaluate the performance of the DPC-CBF algorithm in terms of energy- and time-efficiency, Monte Carlo simulations (500 runs) were performed using the setup described in Section \ref{sec:MC_sim_setup}. Here, the DPC-CBF algorithm was compared against the FIFO and C-CBF benchmarks. The goal is to improve upon the FIFO controller and achieve similar results as the centralized C-CBF controller.

\subsubsection{Comparison with FIFO Control} \label{sec:MCsim_eval_FIFOcomp}
\begin{figure}[!b]
	\centering
	\includegraphics[width=0.85\columnwidth]{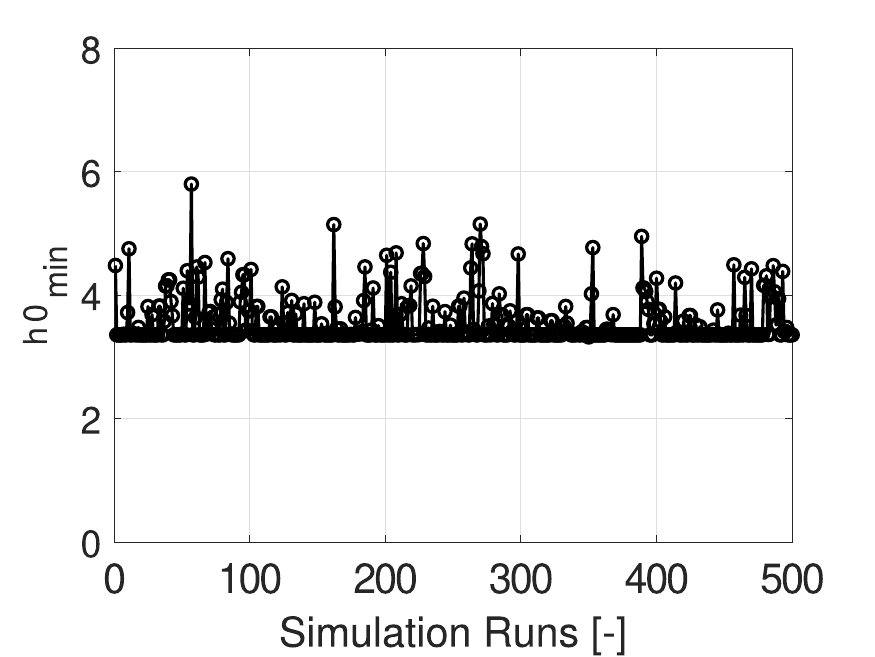}
	\caption{Collision avoidance using DPC-CBF, verified using $h0_{min}$.}
	\label{fig:h0min_DPC-CBF_MCsim}
\end{figure}

The same initial conditions and vehicle masses were used in the DPC-CBF and FIFO cases. Across the 500 runs, no collisions were observed in both the DPC-CBF and FIFO control cases. This is verified by the $h0_{min}$ computed for the DPC-CBF algorithm, illustrated in Fig. \ref{fig:h0min_DPC-CBF_MCsim}. Even though the DPC-CBF runs with no slack variables (i.e., the constraints are hard), there were no infeasibility flags issued by the solver.

\begin{figure}[!t]
	\centering
	\includegraphics[width=\columnwidth]{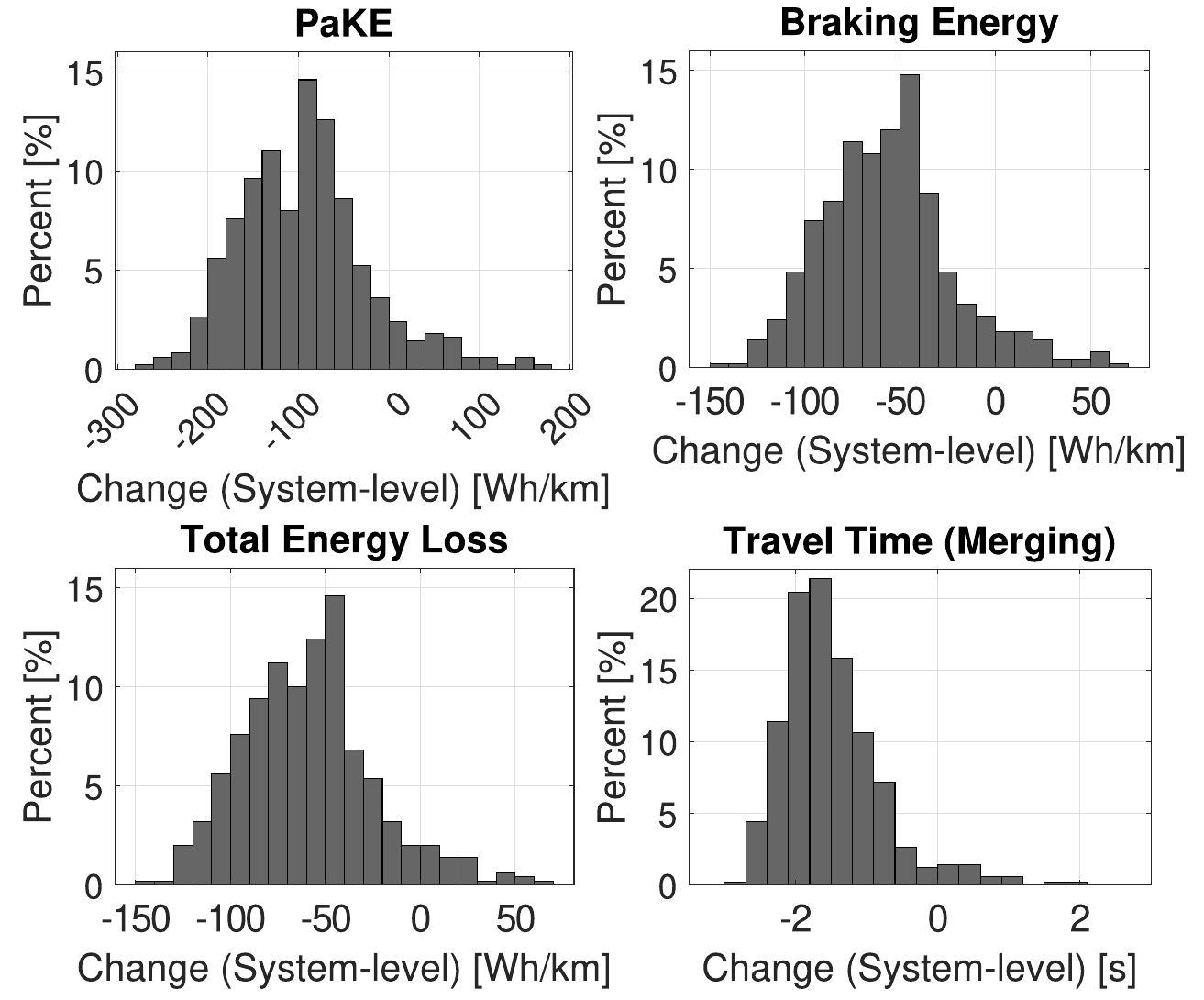}
	\caption{Histograms from Monte Carlo evaluation.}
	\label{fig:MCsim_eval_hist_plots_FIFOcomp}
\end{figure}

Next, the key energy- and flow-related metrics from the Monte Carlo simulations are computed and their distributions are visualized in Fig. \ref{fig:MCsim_eval_hist_plots_FIFOcomp}. Compared with the FIFO benchmark, the DPC-CBF merging controller improves the performance at a system-level across all the evaluation metrics considered. As seen in the DPC-CBF column of Table \ref{table:MCsim_eval_metrics_energy_C-CBF_comp}, the mean PaKE is reduced by $38\%$ while the mean BE, a measure of energy losses from braking, is reduced by over $46\%$. Importantly, the DPC-CBF significantly reduces the mean TEL, accounting for total energy losses, from $251$ Wh/km (FIFO) to $192$ Wh/km, i.e., by around $23\%$. The DPC-CBF controller performed worse than the FIFO control in limited instances where the vehicles were initialized with very symmetric initial conditions.

In addition to the improvement in system-wide energy efficiency, the DPC-CBF approach improves the time efficiency by reducing system congestion. As seen in the DPC-CBF column of Table \ref{table:MCsim_eval_metrics_energy_C-CBF_comp} under the flow-related metrics, the merging time is reduced by $3.5\%$ (average $1.4$ s) while the system-level average vehicle speed of in the defined CZ increases from $20.9$ m/s to $22.1$ m/s i.e., by $5.6\%$.

On analyzing the velocity and acceleration profiles from individual runs, the FIFO approach results in a merging order which is solely determined by the times at which the CAVs enter the CZ. While FIFO schemes consequently lead to orderly (zipper) merge sequences, there are several occurrences when heavier vehicles traveling at higher speeds are forced to brake for lighter and slower ones. This adversely affects the system-wide energy consumption. Such slowdowns propagate as the number of vehicles in the CZ increases, thereby further increasing travel time. In contrast, the DPC-CBF algorithm enables active coordination between the CAVs and the formulation in \eqref{eq:decentralized_PCCA_cost_fn_constr} effectively accounts for heterogeneous traffic. The resulting control policy produces smaller magnitude of velocity changes at a system-level.

\subsubsection{Comparison with C-CBF Control}
While being robust to uncertainties and unexpected behavior, another important feature of the DPC-CBF formulation is that, under nominal conditions, it yields results very close to the centralized  C-CBF solution. This is demonstrated by using Monte Carlo simulations with DPC-CBF and C-CBF algorithms  compared with the same FIFO benchmark.

The DPC-CBF evaluation results discussed in Section \ref{sec:MCsim_eval_FIFOcomp} are retained. Then, for fair comparison, the same simulation setup (including initial conditions and vehicle masses) is used for the C-CBF case. The results from 500 runs of the Monte Carlo simulation are summarized in Table \ref{table:MCsim_eval_metrics_energy_C-CBF_comp}. To ease exposition, only the percent changes in the mean values of the respective evaluation metrics have been tabulated. It is seen that the centralized benchmark performs slightly better than the DPC-CBF algorithm. This behavior is expected -- the C-CBF has exact knowledge of the desired speed of each CAV while the decentralized algorithm uses the current velocity instead while reconciling differences through $\hat w$ predictor-corrector loops.  The minimal performance degradation is offset by reduced V2V communication requirements (only standard BSM needed) and much better robustness to unexpected behavior. 

\begin{table}[!t]
\centering
\caption{Monte Carlo simulations: Comparison between DPC-CBF and C-CBF cases.}
\label{table:MCsim_eval_metrics_energy_C-CBF_comp}
\begin{tabular}{ccc}
\cline{2-3}
 & \multicolumn{2}{c}{\textbf{Percent Change (vs. FIFO) [\%]}} \\ \hline
\textbf{Metric} & \textbf{Benchmark C-CBF} & \textbf{DPC-CBF} \\ \hline
PaKE & $-40.3$  & $-38.0$  \\
BE & $-47.6$ & $-46.6$ \\
TEL & $-23.5$ & $-23.2$ \\ \hline
Travel Time & $-3.6$ & $-3.5$ \\
Average Velocity & $+5.6$ & $+5.5$
\end{tabular}
\end{table}

\section{CONCLUSIONS AND FUTURE WORK}

This paper discusses the decentralized DPC-CBF algorithm for highway merging control of CAVs. The second-order CBF-based safety filter is simultaneously and jointly computing for collision avoidance and prediction of other agents’ actions, adjusting to their motion in stride. The fast predictor-corrector loops in the DPC-CBF controller reconcile differences between predicted and observed motion by other CAVs, while the resulting inter-agent motion avoids gridlocks and unnecessary slowdowns. Benefits over its centralized counterpart, in terms of robustness to unexpected behavior, was presented. Further, Monte Carlo simulations showcased significant improvement in the system-wide energy efficiency and traffic flow compared to a FIFO approach.



\section*{ACKNOWLEDGMENT}

The authors would like to thank Stas Gankov, Mike Brown, Scott Hotz, JoLyn Swain, Gopika Ajaykumar, and Piyush Bhagdikar, all from Southwest Research Institute for contributing to research that led to this paper.



\bibliographystyle{IEEEtran}
\bibliography{IEEEabrv,mybibfile}

\end{document}